\documentstyle{article}
\begin{document}
\vspace*{1.0in}
\begin{center}
Orbital decay of protostellar binaries in molecular clouds \vspace{20pt}\\
U.~Gorti and H.~C.~Bhatt\\
Indian Institute of Astrophysics\\
Bangalore 560034 India \vspace{3.00cm}\\
{\bf Abstract}
\end{center}
The evolution of a protostellar binary system is investigated while
it is embedded in its parent molecular cloud core, and is acted upon
by gas drag due to dynamical friction. Approximate analytical results
are obtained for the energy and angular momentum evolution of the
orbit in the limiting cases, where the velocity is much smaller than,
and much larger than the velocity dispersion of the gas. The general
case is solved numerically. Dissipation causes a decay of the orbit
to smaller separations and orbital eccentricity increases with
time. Binary populations have been statistically generated and evolved
for comparison with observations of the frequency distributions with
period of MS and PMS binaries. Decay of the orbit due to dynamical
friction can circumvent the problem of forming close binaries, as
long-period binaries get dragged and evolve to shorter periods.
\newpage
\begin{center} {\bf 1.~Introduction} \end{center}
Studies of star formation and early stellar evolution
have, in the past, mainly concentrated on the processes involved in the
formation of a single star. These have yielded a plausible sequence of
events that lead to the formation of an optically visible young star from
the gravitational collapse of a dense molecular cloud core (e.g.,
Shu et al 1987). Observations of main-sequence stars and of present-day
star formation sites, however, indicate that binary and multiple systems
of stars are more common and also that most star formation activity
produces groups of stars rather than individual objects. Recent discoveries
of large numbers of pre-main-sequence (PMS) binary systems,
in both high-mass and low-mass star-forming regions, indicate that
binaries are formed very early in the star formation process, probably
even during the collapse phase itself. The formation of binaries thus
appears to involve processes differing from or additional to
those for a single star.

Observational studies of main-sequence binaries of F and G type dwarfs
in the solar neighbourhood (Abt 1983, Duquennoy and Mayor  1991) indicate
certain characteristic features such as a high fraction of stars in
binaries ($\sim$ 70\%), a log--normal frequency--period
distribution with a median at around 180 yr, a  period--eccentricity
relation, and a possible dependance on mass of the secondary mass
distribution. These have provided constraints on theories of binary
formation, which in recent times have been made more stringent by
observations of PMS binaries themselves.
The frequency distribution with period of PMS binaries appears also to
have a single peak, but with seemingly a larger fraction of stars in
binaries, as compared to the MS stars. PMS binaries also exhibit a
wide range of eccentricities and periods and are coeval. For a
recent detailed review of PMS binary stars, refer Mathieu (1994). (Also
see reviews by Reipurth \& Zinnecker 1993, Leinert et al 1993, Bodenheimer
1992 and for theoretical aspects of binary formation, Pringle 1991,
Boss 1992). Theories for binary formation, thus in addition, must also
explain the above observed characteristics of the binary population.
There have been  various mechanisms proposed to form binaries, which
include (a) capture of individual stars by dissipation through tides,
disks or residual gas, (b) fission of a collapsing rotating cloud core
into two nuclei, (c) formation of a binary during hierarchical
fragmentation of a collapsing cloud core and (d) instabilities in
circumstellar disks. The observed smooth distribution of frequency with
period for MS binaries with a single peak, suggests the presence of a
characteristic lengthscale and perhaps a single dominant mode for binary
formation, if there were no further evolution of the binary orbit.
Alternatively, different formation mechanisms acting on
different lengthscales could all be equally important and would be expected
to produce multiple peaks in the frequency distribution of periods.
The observed smooth distribution may then be due to an evolutionary
process subsequent to formation that changes the binary orbital
parameters.

As binaries form as condensations out of a molecular cloud, it
is possible that their general properties depend on the physical
properties of the parent molecular cloud/core. This is especially so,
if the dominant formation mechanism is one of fragmentation, (e.g, Pringle
1991, Bodenheimer et al 1992, Boss 1992) where the
properties of the fragments themselves are largely determined by the
initial conditions in the collapsing cloud (e.g., Bodenheimer 1980, Boss 1992).
In fact, observations of PMS binaries in different star-forming
regions do appear to indicate some differences in their features; the
frequency of binaries may be lower in high-mass star-forming regions
(binary-surveys of the Trapezium cluster region, Prosser et al 1994) and
there appears to be an  excess of short-period binaries
in  the Ophiuchus-Scorpius
region as compared to the Taurus-Auriga region, (Mathieu 1992). Apart from
an intrinsic difference in the nature of the protostellar cores
formed subsequent to fragmentation, these objects probably undergo further
dynamical evolution leading to an evolution of the binary properties as
well. The protostellar binary is formed embedded in relatively massive
amounts of distributed  molecular cloud gas.
Detection of binaries among protostars as young as $10^5$ years,
and the fact that the observed frequency of PMS binaries appears
to be as high as MS binaries, indicates that most binaries are formed
during the early collapse phase. This suggests that a binary remains
embedded in the cloud during much of its initial evolutionary period.
Interactions with the residual gas in the cloud through various effects,
can cause an orbital evolution of the protostellar binary.
In this paper, we study the influence of
gas drag due to dynamical friction (Chandrasekhar 1943)
on the binary, and the time evolution of
its orbital parameters. The effects of introducing the dissipative drag
force are investigated both analytically and numerically.
Dynamical friction may be regarded as arising from a density asymmetry
of the surrounding medium in the wake of a moving object which tends
to retard its motion by exerting a gravitational pull on the object
(Binney \& Tremaine 1987). Drag due to
dynamical friction depends on the density and velocity dispersion of the
ambient medium, and hence orbital evolution under different cloud
conditions is considered. The evolution is also dependent on the binary
parameters, such as the masses of the components, initial separation and
eccentricity of the orbit. Statistically generated samples of binaries with
a range of initial parameters are evolved in time to study changes in gross
features of the binary population due to dissipative drag forces. Approximate
analytical results are presented in Section 2, the numerical methods
are described in Section 3, and Section 4 contains a discussion of the
results obtained. Section 5 discusses a statistical study of PMS binaries.
Sections 6 and 7 contain a discussion on the implications of the results
obtained for binary formation and the main conclusions,
 respectively. \vspace{0.8cm}\\
\begin{center} {\bf 2.~Equations of motion }\end{center}
We consider a uniform density spherical cloud in which the protostellar binary
system is embedded. The forces considered to be acting on each component
of the binary system are (i) the gravity of the cloud, (ii) the gravitational
force due to the other member and (iii) the dynamical friction force due
to the ambient gas. All other forces are neglected. Thus the equations of
motion in a coordinate frame centered on the cloud are , \\
\begin{equation}
{{d^2{\vec{r}_1}}\over{dt^2}}= -4\pi G \rho\vec{r}_1 -
{{G m_2 \vec{r}_{12}}\over{|{\vec{r}_{12}}}|^3} -
4\pi\ln\Lambda_1 G^2 m_1 \rho \left(erf(X_1)-
{{2X_1 e^{-{X_1}^2}}\over{\sqrt{\pi}}}\right){{\vec{v}_1}\over{|\vec{v}_1|^3}}
\end{equation}
\begin{equation}
{{d^2{\vec{r}_2}}\over{dt^2}}= -4\pi G \rho\vec{r}_2 +
{{G m_1 \vec{r}_{12}}\over{|{\vec{r}_{12}}}|^3} -
4\pi\ln\Lambda_2 G^2 m_2 \rho \left(erf(X_2)-
{{2X_2 e^{-{X_2}^2}}\over{\sqrt{\pi}}}\right){{\vec{v}_2}\over{|\vec{v}_2|^3}}
\end{equation}
where the last term in each of the above equations is the dynamical friction
force. Here, $m$ denotes the mass of the protostar, $\rho$
is the density of the cloud and $\ln\Lambda$ denotes the Coulomb logarithm
and is the ratio of maximum to minimum impact parameters. X is a dimensionless
parameter equal to $v/\sqrt{2}\sigma$, where $\sigma$ is the velocity
dispersion of the cloud. The subscripts $1$ and $2$ in each case
refer to the corresponding quantities for the two protostars, and
$\vec{r}_{12}$ the relative position coordinate $\vec{r}_1-\vec{r}_2$. \\
The equations are a set of coupled non-linear differential equations
and an exact solution of which cannot be obtained. Since our interest is
in the evolution of the binary, the orbital evolution can be equivalently
described by the evolution of the angular momentum and energy of the
system. These equations are also not easily solved for the
general case. However they may be
approximately solved in the two  limiting cases of (i) $v \ll \sigma$
where the dynamical friction term is proportional to the velocity $v$
and (ii) $ v \gg \sigma$ when the drag is proportional to $1/v^2$. The
equations can then be described in terms of the reduced mass of the
system. \\
{\em \underline{Case (i): $v \ll \sigma$ }} \\
The velocities of the two masses can be expressed as the sum of the
the relative motion, $v$, and the centre of mass motion $V$. Thus,
\begin{equation}
v_1 = {{m_2 \vec{v}}\over{m_1+m_2}} + \vec{V};\hspace{1.00cm}
v_2 = {{-m_1\vec{v}}\over{m_1+m_2}} + \vec{V}
\end{equation}
{}From equations (1),(2) and (3), with the dynamical friction term now given
by $-\eta m_i \vec{v}_i$ for each particle, we get
\begin{equation}
{{d^2{\vec{r}}}\over{dt^2}}= -4\pi G \rho\vec{r} -
{{G (m_1+m_2) \vec{r}}\over{|{\vec{r}}}|^3} -  \eta {{m_1 m_2}\over{m_1 + m_2}}
 {2 \vec{v}} - \eta (m_1-m_2) \vec{V}
\end{equation}
where $\eta=16\pi^2 G^2  \ln\Lambda \rho/(3 (2 \pi \sigma^2)^{3/2}) $.
 The last term is neglected and the rate of change
of angular momentum can then be written as
\begin{equation}
{{d\vec{L}}\over{dt}}=\mu \vec{r} \times {{d\vec{v}}\over{dt}}=
-2 \mu \eta \vec{L}
\end{equation}
 where $\mu$ is the reduced mass. Therefore,
\begin{equation}
\vec{L} = \vec{L}_0 e^{-2 \mu \eta t}
\end{equation}
The time rate of change of energy and angular momentum of the orbit can
be obtained in a more general manner by rewriting the equations in a
Lagrangian formulation. The dissipative drag force can be expressed as
the velocity gradient of a ``dissipative function'', $\cal{F}$,
(in the limit $v\ll\sigma$) given by
\begin{equation}
{\cal{F}} =   \eta \mu^2 v^2
\end{equation}
The rate of change of angular momentum is obtained from one of
the Lagrange equations, and is equal to $-2 \mu \eta L$, as was already
obtained in equation(5).
The rate of change of energy is given by (refer for e.g, Goldstein 1977)
\begin{equation}
{{dE}\over{dt}}= - \sum{{{\partial{{\cal{F}}}}\over{\partial{\dot{q_j}}}}
\dot{q_j}} = - 2 \eta \mu^2 v^2
\end{equation}
where $\dot{q_j}$ represent the generalized velocities.
It can be assumed that the system evolves
quasi-virially, such that the virial theorem always holds during the
motion. The total energy, $E$, is thus equal to one-half the kinetic
energy and for a circular orbit, the velocity can be expressed in
terms of the total energy as $v^2 = - 2 E/\mu$. Thus,
\begin{equation}
{{dE}\over{dt}}= 4 \eta \mu E \hspace{1.00cm}{\rm and}\hspace{1.00cm}
E=E_0 e^{4 \eta \mu t}
\end{equation}
The equations have also been numerically solved (see Section 3) and
a comparison with the analytical solutions (equations (5) and (8))
shows that though the equations have been solved for the circular
orbit, they approximate reasonably well for higher-eccentricity orbits.
The rate of change of eccentricity with time cannot, however, be
obtained from the present solutions, as the orbit always remains
circular. Figure 1 shows the time evolution
of angular momentum, energy, and semi-major axis of the binary system.\\ \\
{\em \underline{Case (ii): $v \gg \sigma$ }} \\
For velocities large compared to the local velocity dispersion of the
background, the dynamical friction force is proportional to $1/v^2$ and
the equations of motion are
\begin{equation}
{{d^2{\vec{r}}}\over{dt^2}}= -4\pi G \rho\vec{r} -
{{G (m_1+m_2) \vec{r}}\over{|{\vec{r}}}|^3} - \tilde{\eta} \left(
{{m_1\vec{v}_1}\over{|\vec{v}_1|^3}} -
{{m_2\vec{v}_2}\over{|\vec{v}_2|^3}} \right),
\end{equation}
where $\tilde{\eta} = 4 \pi \ln\Lambda G^2 \rho$.
For simplicity, it is assumed that the two masses are equal and
${v_1}^3={v_2}^3\approx(v^2/4 + V^2)^{3/2}$ so that equation(10)
reduces to
\begin{equation}
{{d^2{\vec{r}}}\over{dt^2}}= -4\pi G \rho\vec{r} -
{{2 G m \vec{r}}\over{|{\vec{r}}}|^3} -
{{ \tilde{\eta} m \vec{v}}\over{(v^2/4 + V^2)^{3/2}}}
\end{equation}
where m represents the mass of the protostars. The dissipative function
${\cal{F}}$ is thus
\begin{equation}
{\cal{F}} = - {{2 \tilde{\eta} m^2} \over{(v^2/4 +  V^{\ 2})^{1/2}}}
\end{equation}
The rate of change of
energy is again given by equation(8) and is integrated by assuming
that the velocity of the center of mass is a constant, and that virial
equilibrium always holds. Therefore, for a circular orbit,
\begin{equation}
\frac{2}{3}\sqrt{E_k - E}(4E_k -E) + {E_k}^{3/2}
\ln{{\sqrt{E_k - E}-\sqrt{E_k}}\over{\sqrt{E_k - E}+\sqrt{E_k}}} =
2 \eta m^{5/2} t + c
\end{equation}
where $E_k$ is equal to $m V^{\ 2}$ and c is the constant of integration, to
be determined from the initial conditions. The equation of motion
for a particle orbiting a
central point mass was studied by Hoffer (1985) who developed approximate
expressions for the evolution of the semi-major axis of the orbit. It can
be easily shown that the above equation  corresponds to Hoffer's solution
in the limit of vanishing $E_k$. The angular momentum equation can be
obtained as above
\begin{equation}
{{d\vec{L}}\over{dt}}=-{{\tilde{\eta} m \vec{L}}\over{(v^2/4+V^{\ 2})^{3/2}}}
\end{equation}
This can be integrated for a circular orbit to give the evolution of
angular momentum
\begin{equation}
\left(-{{4 V^{\ 2}}\over{3}} -{{k^2}\over{12 L^2}}\right)
\left({{k^2}\over{4 L^2}}+V^{\ 2}\right)^{1/2} +
V^{\ 3} \ln\left[V^{\ 2} L + L V \left({{k^2}\over{4 L^2}}+V^{\ 2}
\right)^{1/2}\right]
= -\tilde{\eta} m t + c
\end{equation}
where $k$ is equal to $G m^2$ and c is a constant to be determined from the
initial conditions. The equations (13) and (15) are transcendental in nature
and the energy and angular momentum cannot be easily expressed as explicit
functions of time. The nature of the evolution can, however, be inferred.
Initially, both energy and angular momentum are exponential functions
of time  with a timescale of the order $2 \tilde{\eta} m/ V^{\ 3}$ and
$\tilde{\eta} m/ V^{\ 3}$ respectively. For typical cloud conditions
surrounding
a protostar, with densities $\sim 10^6$ cm$^{-3}$ (e.g., $\rho$ Oph),
these timescales are of the order of a few $10^5$ years. As the binary
loses kinetic energy, the protostars fall in and become more tightly
bound. The relative velocity of the binary thereby increases and change in
energy and angular momentum of the binary is no longer exponential, but
slower. At later times (when the relative velocity $v \ll V$, the center
of mass velocity), the energy varies as $ \approx t^{2/3}$ and
angular momentum as $\approx t^{1/3}$.
Figure 2 shows the variation of energy, angular momentum and semi-major
axis of the binary system with time.
\vspace{0.80cm}\\
\begin{center} {\bf 3.~Numerical Solutions }\end{center}
The equations governing the motion of the binary can be solved only in
the limiting cases, where the dynamical friction term reduces to a simple
form, and by making certain approximations.
To study the complete evolution of the binary system, and to check the
validity of the above derived expressions, equations(1) and (2) have been
solved exactly using numerical methods. Also, the above expressions do not
yield the variation of eccentricity with time, as it has been assumed that
the orbit is always circular. The eccentricity of the orbit increases or
decreases with time, depending on the relative rates of change of energy
and angular momentum. Though dissipative forces are in general believed
to circularise orbits, this is not necessarily so, one such exception being
the drag due to dynamical friction (e.g., Hoffer 1985). The numerical
solutions allow a study of the eccentricity variation with time and also
a more exact investigation of the other orbital parameters.

The problem essentially involves two very different timescales, one
being the period of the binary, which may be of the order of years, and
the other the drag timescale which can be estimated to be of the order
of  $\sim 10^{5-6}$ years, for usual protostellar cloud parameters.
Thus the orbit has to be typically integrated over a few $10^6$ periods
and to meet requirements of accuracy, forces are evaluated about a
thousand times each orbit. In order to check the accuracy of the
integration scheme to be adopted, the binary is first evolved in the
absence of drag for about $10^6$ years. Ideally, energy conservation
is to be expected, which is not in general acheived by standard methods
of integration that use a variable time-step.
Ensuring energy conservation by choosing a constant time-step
criterion (and thus a time-symmetric integrator) turns out to be prohibitively
expensive and we adopt the presription by Hut, Makino and McMillan (1995)
for a time-symmetrized leapfrog with a variable time-step. The second order
method has been chosen over the fourth-order Hermite integrator as the
latter has proved less economical for the present problem. The symmetrized
leapfrog integrator involves the use of an implicit time-step criterion
which preserves the time-symmetry of the original equations and excellent
energy conservation (with the exclusion of drag) is obtained.
The orbits are then integrated including the drag term in the forces.

Figure 3 shows the variation in time of energy, angular momentum,
eccentricity and semi-major axis. Also shown superposed on these plots
are the corresponding analytical solutions (for the case $v \gg \sigma$).
It can be seen that there is reasonable agreement between the two,  even
though the orbit is not circular.
\vspace{0.80cm} \\
\begin{center} {4.~\bf Evolution of a binary in a cloud}\end{center}
As seen above, the predominant effect of the drag force on the binary is
to shrink the orbit to smaller separations. The extent to which the
orbit shrinks in a given period of time depends on the initial
separation of the binary components. The initial semi-major axis of the
binary determines the initial relative velocity, and hence the magnitude
of the drag term in the force initially. As the masses lose energy and
angular momentum and spiral in, their orbital separation decreases and
there is an increase in the magnitude of their relative velocity. If the
initial velocity is less than the gas velocity dispersion $\sigma$, then
the drag force increases in magnitude in the beginning as $v$ and
decreases later as $1/v^2$. Thus, there is an initial rapid
decay of the orbit and as the relative velocity increases, drag
decreases and the decay is slower.
 Figure 4 shows the evolution of semi-major
axis for different initial separations during a fixed interval of time.
The cloud in which the protostellar binary is embedded is assumed to be
of constant number density ($10^6$ cm$^{-3}$) and with a velocity dispersion
of 2.3 $kms^{-1}$. These parameters are typical of the environment of
protostars and have been chosen to comply with observations of the
prototypical star-forming cloud core, $\rho$ Ophiuchus. It is seen that
the timescale for a decrease in the semi-major axis by a factor of
two, $t_{1/2}$, is about a few $10^5$ years.

A distinct feature of drag due to dynamical friction, is
that orbits get increasingly more eccentric with time. This implies that the
fractional loss of angular momentum exceeds that of energy.  The eccentricity
increases with time and gradually tends towards one.
We have studied eccentricity variations for differing initial
eccentricity and Figure 5 shows the initial and final eccentricity
for an orbit with the same initial separation, after $10^6$ years.
Also shown are the changes in angular momentum and energy
with time. As is to be expected, loss of angular momentum )
is greater for orbits with higher eccentricity.
The increase of eccentricity with time indicates that
binary systems that have evolved in the presence of gas and hence
suffered drag due to it, must lie on high--eccentricity orbits in
general. While this may be so for orbits with long periods and hence
larger separations, it need not hold for the shorter period binary
systems. It is believed that almost all young protostars are accompanied
by massive circumstellar disks, with typical extents of a $\sim$ 100 AU,
(Strom et al 1993). Therefore, for separations much shorter than these,
the presence of the disks has to be taken into account for a more
realistic analysis. It is not clear how dissipation caused by disks
affects the eccentricity of the orbit, though at very short periods,
tidal forces may rapidly tend to make the orbits more circular (Clarke
1992). As circumstellar disks have not been included in our present
analysis, we have not considered the evolution of short-period binaries,
as we believe that the effects of disks on the binary orbit is probably
more prominent at these separations.

The effects of variation in cloud gas density can be assessed easily
from equations(8),(10) and (11). The relevant timescales vary inversely
with density, and the evolution of semi-major axis for three different
cloud densities is shown in Figure 6. The evolution for different
gas velocity dispersions  depends on the velocity of the binary and
hence on the separation. Figure 6 also shows the semi-major axis plotted
against time for different velocity dispersions. The cloud density
has been kept constant at $\sim 10^6 cm^{-3}$. The drag force depends
sensitively on the velocity dispersion, and is greater for lower velocity
dispersions. This implies that evolution of binaries subsequent to
formation may, in general, be different for different clouds (with
differing velocity dispersions or differing densities). There seems to
be tentative observational evidence for differences in the frequency of
short-period binaries in $\rho$ Ophiuchus and Taurus-Auriga clouds. If the
observed excess of closer binaries in $\rho$ Ophiuchus is not because of
a larger incidence of formation, evolution of PMS binaries in the presence
of drag could be a possible explanation. $\rho$ Ophiuchus has a lower
velocity dispersion and a higher average gas density as compared to the
Taurus-Auriga star forming region, and hence a drag force of greater
magnitude. This implies that evolution to shorter orbital separations
is faster in $\rho$ Ophiuchus and it is more likely to contain short-period
PMS binaries.

It is to be noted that though only the dissipative effects of dynamical
friction have been considered here, in principle, other forces could
also drag the orbital evolution. For close binaries, disks and tidal
torques provide active dissipative agents in addition to dynamical friction
and are probably more important. Ram pressure due to the surrounding gas
is another potential source of drag, and for self-gravitating objects
like protostars, is similar in form to the dynamical friction term (in the
limits of large velocity), except for being a factor of $\ln\Lambda$ lower
in magnitude. The results derived here therefore apply equally well for
ram pressure deceleration as well, but on longer timescales. \vspace{0.80cm}\\
\begin{center} {\bf 5.~Statistical studies of PMS binaries}\end{center}
High-resolution techniques and better infra-red detectors have in
recent times enabled the detection of large numbers of PMS binaries
covering a wide range of periods. A comparison with observations of
main-sequence (MS) binaries can thus be made and it is found that the
fraction of PMS stars in binary systems is equal to or even larger than
the frequency of MS binaries ($\sim 60-70\%$). The periods of MS binaries
form a smooth distribution in the range 0 $<$ Log P(days)$ <$ 9, with a
maximum at around 180 years. The eccentricities span a wide range and
only the shortest period orbits are circular.  The masses of the primary
and secondary are probably uncorrelated and are consistent with a random
pairing from an initial mass function (Duquennoy \& Mayor 1991).
Periods of PMS binaries within the range observed appear to follow a
similar distribution as that of the MS binaries. PMS binary orbits also have
a range of eccentricities. Since drag affects the orbital evolution
of a binary system as discussed above, it is to be expected that this will
be manifest in the observed properties of PMS and MS binaries. In order to
investigate the effects of gas drag on the class of PMS binaries as a
whole, statistical samples of binary systems are generated and evolved
in time. The initial conditions and assumptions made are as follows:
\begin{enumerate}
\item The masses are determined from a Salpeter initial mass function
($dN/dM \propto M^{-2.35}$) within the range 0.5 $M_{\odot}$ to
5 $M_{\odot}$. Binaries are formed by randomly pairing two masses.
\item The eccentricities are random and constrained within
the interval 0-0.8. Higher eccentricity orbits were not considered
due to difficulties in maintaining the accuracy during numerical
integration.
\item The separations (and hence periods) of the entire sample follow
an assigned distribution. Two such initial distributions were considered,
one where the number of binaries {\em per separation interval} are equal
and another where the number of binaries are equal {\em per logarithmic
separation interval}.
\item The separations (semi-major axes) are also randomly assigned
to the binaries and lie between $100$ AU and $10^4$ AU. The lower
limit has been somewhat arbitrarily chosen and corresponds to the
typical size of a protostellar disk, beyond which the disk is expected
to  significantly influence the orbital evolution. The upper limit
corresponds to an average expected separation for protostars in a
clustered environment, beyond which the binary may not remain bound.
\item The binary is given a random position in the uniform density
cloud (of density $10^6 cm^{-3}$ and velocity dispersion $2.3 kms^{-1}$)
and assigned the local virial velocity with an arbitrary direction.
\end{enumerate}
The sample consists of 100 binaries for the case with an equal
number per separation interval (Case I) and 70 binaries for the  case with
equal number of binaries per logarithmic separation interval (Case II).
Each binary orbit is integrated in time for about $5\ 10^6$ years.\\
The frequency distribution as a function of period at different instances
of time for Case I is shown in Figure 7.
The initial distribution is  constant with separation, as shown in Panel (a).
As the drag force varies as the inverse square of the velocity, binaries with
larger separations (and smaller velocities) suffer greater drag and lose
energy at a faster rate as compared to the higher velocity binaries at
shorter periods. The frequency distribution gradually gets altered and
a peak appears, which gets more pronounced at later times. The position
of the peak slowly shifts towards shorter periods if the evolution is
allowed to proceed for longer times.
For an initial distribution that is constant in equal logarithmic
intervals of separation, the frequency distribution also gets
gradually altered as shown in Figure 8. A distinct peak in the frequency
distribution appears at later times than in Case I probably because
of the presence of a comparatively larger number of shorter period binaries
which evolve more slowly. \\
The observed maximum in the frequency distribution with period can
perhaps possibly be explained as an effect of gas drag (or other
disipative forces) which determines the initial evolution of a binary
while still embedded in the molecular cloud. The position of the
maximum then depends mainly on the cloud parameters and to a much
lesser extent on the duration of the embedded phase of a typical
binary system. The observed peak in the MS binary distribution would
then be representative of the typical or average cloud conditions
in the Galactic star-forming environment. \\
The eccentricity of a binary acted upon by dynamical friction
increases with time. This implies that binary orbits that have
decayed to shorter periods are also more eccentric. Observations of
MS binaries seem to show the reverse trend, with close binaries
lying on nearly circular orbits. It is believed that for very close
binaries, tidal forces rapidly tend to circularize the orbits (e.g.,
Mathieu 1994). The trend towards low eccentricities is however seen only
for orbits with periods shorter than $\sim$ few 100 days, and our analysis
does not extend to such short-period binaries.
Figure 9 shows the observed eccentricities
of MS binaries and those from the above samples,  plotted against
the period of the orbit for the overlapping range of
periods ( 3.5 $<$ Log P (days) $<$ 5).
It is seen that within the range of periods observed, there is no
contradiction between the results obtained here and observations.
\vspace{0.80cm}\\
\begin{center} {\bf 6.~Implications for binary formation}\end{center}
The presence of a peak in the frequency distribution with separation
for MS and PMS binaries suggests the existence of a preferred
lengthscale for the formation of binaries. Existing mechanisms
for binary formation include fragmentation, fission, capture processes
and instabilities in disks. Of these, fragmentation is  currently
believed to be the primary formation process (see reviews by
Bodenheimer 1992, Pringle 1991, Boss 1992) where the protostar
fragments during the isothermal collapse phase or alternatively where
the binary is formed as a result of a hierarchical fragmentation
process. The frequency distribution arising from fragmentation is
not well-determined (Boss 1992), but formation of binaries with
larger separations is easier or more favoured than the formation
of close binaries (Bodenheimer 1992). Decay of the orbit due to
dynamical friction can circumvent the problem of forming close binaries,
as long-period binaries get dragged and evolve to shorter periods. \\
If all (or any other) of the above mentioned mechanisms are equally
important for the formation of binaries, there are intrinsically
different lengthscales involved in each case. (E.g., the
separation of a binary cannot be much larger than the extent
of the disk if the binary forms by gravitational instabilities
in a disk). It is then very unlikely for the initial frequency
distribution to be a uniform function of period. In such a case,
dissipation of energy due to drag may serve to smoothen out the
irregularities, though the precise nature of the resultant
frequency distribution would depend on the initial
distribution. \vspace{0.8cm} \\
\begin{center} {\bf 7.~Conclusions}\end{center}
The evolution of a protostellar binary system is investigated while
it is embedded in its parent molecular cloud core, and is acted upon
by gas drag due to dynamical friction. Approximate analytical results
are obtained for the energy and angular momentum evolution of the
orbit in the limiting cases, where the velocity is much smaller than,
and much larger than the velocity dispersion of the gas. The general
case is solved numerically. Dissipation causes a decay of the orbit
to smaller separations and orbital eccentricity increases with
time. For a protostellar binary embedded in a typical star-forming cloud
core (of density $\sim 10^6$ cm$^{-3}$ and velocity dispersion 2.3 kms$^{-1}$)
the decay time for the semi-major axis of the orbit to half its
initial value is about $10^{5-6}$ years. As a binary probably spends about
a few $10^6$ years embedded in the cloud, significant evolution of the
orbit is expected before the dispersal of the cloud.

Binary populations have been statistically generated and evolved
for comparison with observations of MS and PMS binaries.
Dynamical friction changes the initial frequency distribution with period
of the binaries, and a maximum is obtained in the distribution as
seen observationally. Two different initial distributions are considered,
and the peak is found to shift slowly  towards shorter periods at later
times.
Decay of the orbit due to gas drag causes an evolution of long-period orbits
to shorter periods and thus enables the formation of close binaries.
\vspace{0.8cm} \\
\begin{center} {\bf References} \end{center}
Abt, H.A., 1983, Ann. Rev. of Astron. Astrophys., 21, 343 \\
Binney, J., Tremaine S., 1987, In ``Galactic dynamics'', Princeton
  University Press, Princeton, p.791 \\
Bodenheimer, P., 1992, In ``Star formation in Stellar systems'', Cambridge
 University Press, Cambridge, p.1\\
Bodenheimer P., Tohline J.E., Black D.C., 1980, Astrophys.J, 242, 209 \\
Boss A.P., 1992, In ``Complementary approaches to double and multiple
  star research'', IAU Coll. 135, p.195 \\
Chandrasekhar S., 1943, Astrophys.J, 97, 255 \\
Clarke C., 1992, In  ``Complementary approaches to double and multiple
  star research'', IAU Coll. 135, p.176 \\
Duquennoy A., Mayor M. 1991, Astron. Astrophys., 248, 485\\
Goldstein, H. 1977, In ``Classical Mechanics'', p.24 \\
Hoffer, J., 1985, Astrophys. J.,  289, 193 \\
Hut P., Makino J., McMillan S., 1995, Astrophys. J. Lett., 443, L93 \\
Lienert Ch., Weitzel N., Zinnecker H., Christou J, Ridgeway S, et al
 1993, Astron. Astrophys., 278, 129 \\
Mathieu, R.D., 1992, In ``Complementary approaches to double and multiple
 star research'', IAU Coll. 135, p.30\\
Mathieu, R.D., 1994, Ann Rev. of Astron. Astrophys., 32, 465 \\
Pringle J., 1991, In ``The physics of star formation and early stellar
  evolution'', Kluwer, Dordrecht, p.437 \\
Prosser C.F., Stauffer J.R., Hartmann L., Soderblom D.R., Jones B.F.,
   Werner M.W., McCaughrean M.J., 1994, Astrophys.J., 421, 517 \\
Reipurth B., Zinnecker H., 1993, Astron. Astrophys., 278, 81 \\
Shu, F.H., Adams F.C., Lizano, S. 1987, Ann. Rev. of Astron. Astrophys, 25,
23 \\
Strom S.E., Edwards S., Strutskie M., 1992, In ``Protostars and Planets III'',
eds Levy E.H., Lunine J.I., Tucson:University of Arizona Press, p837\\
\newpage
\begin{center}
{\bf Figure Captions:} \vspace{0.80cm} \\
\end{center}
\begin{itemize}
\item[Figure 1] The energy, semi-major axis and angular momentum are plotted
against
time, scaled by their corresponding initial values. (Case 1: $ v \ll \sigma$)
\item[Figure 2] The energy, semi-major axis and angular momentum are plotted
against
time, scaled by their corresponding initial values. (Case 2: $ v \gg \sigma$)
\item[Figure 3] The energy, semi-major axis, angular momentum and eccentricity
are plotted againts time. The solid lines correspond to the numerical result
and
the dashed lines to the analytical approximation.
\item[Figure 4] Evolution of semi-major axis for different initial separations.
Time is plotted in units of $10^6$ years.
\item[Figure 5] For different initial eccentricities of an orbit with  the same
initial separation,  the eccentricity, energy and angular momentum
after a fixed interval of time are shown.
\item[Figure 6] Evolution of semi-major axis of a binary for different cloud
densities (Panel (a)) and velocity dispersion of gas in the cloud (Panel (b)).
Time is in units of $10^6$ years.
\item[Figure 7] Frequency distribution with period of binaries which are
initially distributed equally in equal intervals of separation. (Case I)
\item[Figure 8] Frequency distribution with period of binaries which are
initially distributed equally in equal  logarithmic intervals of
 separation. (Case II)
\item[Figure 9] The eccentricity-period distribution for MS binaries (open
circles) (from Duquennoy \& Mayor 1991) and PMs binaries after $1.8\ 10^6$
 years
(filled circles). (Plotted for Case I).
\end{itemize}
\end{document}